\documentclass[aps,prl,reprint,groupedaddress,amsmath]{revtex4-1}
\usepackage{graphicx}
\usepackage{amsmath}

\begin{document}

\title{Tests of universal three-body physics in an ultracold Bose-Fermi mixture}

\author{Ruth S. Bloom}
\author{Ming-Guang Hu}
\author{Tyler D. Cumby}
\author{Deborah S. Jin}

\affiliation{JILA, NIST and University of Colorado; Department of Physics, University of Colorado, Boulder}

\date{\today}

\begin{abstract}
Recent measurements of Efimov resonances for a number of ultracold atom species have revealed an unexpected universality, in which three-body scattering properties are determined by the van der Waals length of the two-body interaction potential. To investigate whether this universality extends to heteronuclear mixtures, we measure loss rate coefficients in an ultracold trapped gas of $^{40}$K and $^{87}$Rb atoms.  We find an Efimov-like resonance in the rate of inelastic collisions between $^{40}$K$^{87}$Rb Feshbach molecules and $^{87}$Rb atoms.  However, we do not observe any Efimov-related resonances in the rates of inelastic collisions between three atoms. These observations are compared to previous measurements by the LENS group of Efimov resonances in a $^{41}$K and $^{87}$Rb mixture as well as to recent predictions.
\end{abstract}


\maketitle
Magnetically tunable Feshbach resonances in ultracold atomic gases make it possible to explore a vast array of phenomena~\cite{ChinRMP}, and recent years have witnessed a rapid evolution in the understanding of universal few-body effects that accompany resonantly enhanced two-body interactions.  First proposed in the context of nuclear theory, the Efimov effect consists of an infinite series of three-body bound states when at least two of the three pairwise interactions have large two-body scattering length $a$~\cite{Efimov70,Efimov73,Braaten06,Zinner13}. These Efimov trimers exist even when the two-body potentials do not support a bound state, and their spatial extent reaches far beyond the scale of inter-particle forces. Notably, Efimov trimers follow a discrete scale invariance, with each subsequent bound state larger than the last by a scaling factor of $e^{\pi/s_\text{0}}$, where $s_\text{0}$ depends on the number of resonant interactions as well as the quantum statistics and the mass ratio of a trimer's constituents~\cite{Braaten06,DIncao06}.

Several ultracold atom experiments have now leveraged tunable interactions to measure Efimov physics ~\cite{Kraemer06,Ottenstein08,Huckans09,Zaccanti09,Barontini09,Pollack09,Nakajima10,Gross10,Berninger11,Wild12}. In dilute gases, weakly bound Efimov trimer states mediate inelastic collisions and give rise to a rich spectrum of atom loss resonances as a function of $a$~\cite{Braaten06}.  A surprising pattern emerged in homonuclear systems, where Efimov loss resonances are consistently measured at $a_-$$\approx$$-9.1~R_\text{vdW}$ for different nuclear spin states~\cite{Gross10}, different Feshbach resonances~\cite{Berninger11}, and different atomic species~\cite{Ottenstein08,Huckans09,Zaccanti09,Pollack09, Gross10, Berninger11,Wild12}, where $R_\text{vdW}$ is the two-body van der Waals length. The three-body parameter that sets the locations of Efimov resonances was previously understood to be a non-universal quantity subject to each system's unique interaction potential at short range~\cite{Efimov73,Braaten06,DIncao09}. However, recent theoretical work has shown that van der Waals potentials of cold atoms screen these non-universal details~\cite{Chin11,Naidon12,WangJ12,Schmidt12,Sorensen12}. A universal three-body parameter has also been predicted for heteronuclear mixtures~\cite{WangY12b}.

Extending the experimental study of Efimov trimers to include mixtures of species is of particular interest to universal few-body physics~\cite{DIncao06,Marcelis08,Barontini09,Helfrich10,WangY12a}. Halo nuclei, in which two neutrons orbit a heavier nuclear core at a distance beyond the classically allowed radius, are an example of Efimov-like states with mass imbalance~\cite{Jensen04,Frederico12}.  To date, the only measurement of a mixed-atom-species Efimov trimer was done by the LENS group for a gas of $^{41}$K and $^{87}$Rb atoms~\cite{Barontini09}.  At JILA, a previous study of inelastic loss in a different isotopic mixture, $^{40}$K and $^{87}$Rb, found no Efimov features~\cite{Zirbel08}. However, given recent predictions for this system~\cite{Helfrich10,WangY12b}, we now revisit the $^{40}$K-$^{87}$Rb mixture with additional loss measurements aimed at exploring the possibility of a universal three-body parameter for heteronuclear mixtures.

Our measurements start with an ultracold mixture of bosonic $^{87}$Rb atoms in the $|f,m_f\rangle$=$|1,1\rangle$ state and fermionic $^{40}$K atoms in the $|9/2,-9/2\rangle$ state, which are the lowest energy hyperfine states in a magnetic field.  Here, $f$ corresponds to the atomic angular momentum and $m_f$ is its projection. The atom gas is prepared near the onset of quantum degeneracy, following methods outlined in~\cite{Cumby13}. The temperature $T$ is above $1.1~T_\text{c}$ for $^{87}$Rb and above $0.2~T_\text{F}$ for $^{40}$K, where $T_\text{c}$ is the transition temperature for Bose-Einstein condensation and $T_\text{F}$ is the Fermi temperature.  Both atom species are imaged a few milliseconds after their release from a single-beam optical trap. In-trap density profiles are determined from fits to Gaussian distributions, assuming ballistic expansion. An $s$-wave magnetic Feshbach resonance is used to control the interactions between $^{87}$Rb and $^{40}$K atoms, where $a$ at a magnetic field $B$ is given by $a$=$a_\text{bg}(1$-$\frac{\Delta}{B-B_\text{0}})$, $a_\text{bg}$=$-187~a_\text{0}$, $B_\text{0}$=$546.618$~G, $\Delta$=$-3.04$~G, and $a_\text{0}$ is the Bohr radius~\cite{Klempt08}.

We begin by considering inelastic atom-molecule collisions, which trigger the vibrational relaxation of weakly bound Feshbach molecules into deeply bound states.  The energy released by this process ejects both collision partners from the trap.  The inelastic collision rate is enhanced when $a$ is tuned to a value, denoted $a_*$, where an Efimov trimer state is degenerate with the atom-molecule threshold~\cite{Nielsen02,Braaten06}.  Such Efimov resonances have been observed as broad loss features for $^{133}$Cs~\cite{Knoop09} and $^{6}$Li dimers~\cite{Lompe10,Nakajima10}.

 \begin{figure}
 \includegraphics[width=0.5\textwidth]{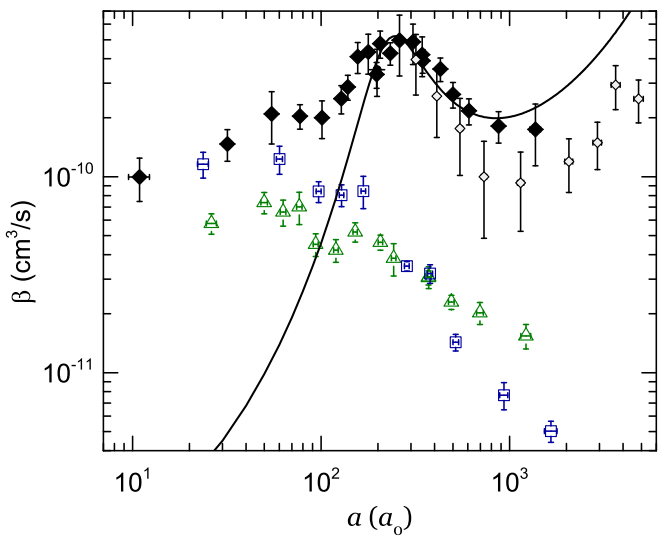}
 \caption{\label{fig:beta}(Color online) Measured $\beta$ vs.~$a$ for collisions of $^{40}$K$^{87}$Rb molecules with $^{40}$K atoms in the $|9/2,-7/2\rangle$ state (triangles), $^{40}$K atoms in the $|9/2,-9/2\rangle$ state (squares), and $^{87}$Rb atoms in the $|1,1\rangle$ state (solid diamonds). Vertical error bars indicate one standard deviation of statistical error, and horizontal error bars correspond to an upper bound of $\pm0.02$~G variation in $B$ across the cloud. A broad resonance is evident in the $^{87}$Rb+$^{40}$K$^{87}$Rb loss and the solid curve shows a fit to Eq.~(\ref{eqn:betashape}).  A previous measurement of $^{87}$Rb+$^{87}$Rb$^{40}$K loss at lower temperature~\cite{Zirbel08} (open diamonds) agrees well with our data, but did not cover the broad resonance. In plotting the previous data, we adjust $a$ to match a more recent calibration of the Feshbach resonance parameters that is used throughout the present work.}
 \end{figure}

For collisions of $^{87}$Rb atoms with $^{40}$K$^{87}$Rb Feshbach molecules, Helfrich~$et~al.$ have suggested that an Efimov resonance exists outside the range of previous measurements~\cite{Helfrich10,Zirbel08}.   Here, we extend these data to lower values of $a$. Similar to the previous work~\cite{Zirbel08}, we measure the molecule loss for three different collision partners: $^{87}$Rb atoms in the $|1,1\rangle$ state, $^{40}$K atoms in the $|9/2,-9/2\rangle$ state, and $^{40}$K atoms in the $|9/2,-7/2\rangle$ state.  Efimov resonances are only possible for the first case, and, in the last case, the collision partner does not participate in the Feshbach resonance.

For these measurements, the atom gas mixture is typically prepared at $a$=$-300~a_\text{0}$ and $T$=$300$~nK. The magnetic field is slowly swept across the Feshbach resonance at a rate of 3 G/ms to form $^{40}$K$^{87}$Rb Feshbach molecules by magnetoassociation~\cite{Cumby13}.  To prepare a pure sample of molecules and one particular atom species, we use a combination of resonant light and RF pulses to selectively remove unpaired atoms from the trap. When measuring non-resonant collisions, we also transfer the unpaired $^{40}$K atoms into the $|9/2,-7/2\rangle$ hyperfine state at this point.  The $^{87}$Rb atom removal is most efficient for a small sample, so $^{40}$K+$^{40}$K$^{87}$Rb collisions are measured with an additional confining laser beam that increases the $^{40}$K axial trap frequency to 30~Hz.

To measure trap loss rates, we sweep the magnetic field to the value desired for the loss measurement at a rate that is fast compared to loss, wait a variable hold time $t$, turn off the optical trap, and then jump the magnetic field back across the Feshbach resonance at a rate of 50 G/ms to dissociate the molecules. We extract the number of molecules, the number of free atoms, and the in-trap cloud sizes from images of the resulting atom clouds~\footnote{In-trap radial sizes are calculated from the measured axial size and the trap aspect ratio}. Using pure molecule samples, we find molecule-molecule loss to be negligible. The atom-molecule collisions can then be characterized by a loss rate coefficient $\beta$ defined by
\begin{equation}
 \dot{N}_\text{KRb}(t)=\dot{N}_\text{A}(t)=-\beta \int d^3\textbf{r}~n_\text{KRb}(\textbf{r},t)~n_\text{A}(\textbf{r},t),
 \label{eqn:beta}
\end{equation}
where $n_\text{KRb}(\textbf{r},t)$ is the number density of the molecules and $n_\text{A}(\textbf{r},t)$ is the number density of the atoms.  We extract $\beta$ by fitting the measured loss of atoms and molecules to numerical solutions of Eqs.~(\ref{eqn:beta}). These solutions account for heating by including the measured cloud sizes at each point in time.  The estimated systematic uncertainty in $\beta$, which is dominated by the uncertainty in the in-trap densities, is $\pm20\%$.

Figure \ref{fig:beta} plots the measured $\beta$ vs.~$a$.  The rate of molecule collisions with $^{40}$K in either hyperfine state decreases smoothly with increasing $a$, as was seen previously~\cite{Zirbel08}.  In contrast, for collisions between $^{40}$K$^{87}$Rb molecules and $^{87}$Rb atoms, we observe a prominent resonance whose location lies beyond the range of previous data~\cite{Zirbel08} and is consistent with the suggestion of $a_*$=$200(50)~a_\text{0}$ made by Ref.~\cite{Helfrich10}. The universal shape of an atom-molecule Efimov resonance at $T$=0 is~\cite{Helfrich10}
\begin{equation}
 \beta(a)=C_\beta\frac{\text{sinh}(2\eta_*)}{\text{sin}^2[s_0~\text{ln}(a/a_*)]+\text{sinh}^2(\eta_*)}\frac{\hbar a}{m_\text{1}}.
 \label{eqn:betashape}
\end{equation}
Here, $\eta_*$ sets the Efimov resonance width and $a_*$ sets the scattering length at the peak of the Efimov resonance. In the $^{40}$K-$^{87}$Rb mixture, $s_0$=0.6536, $m_\text{1}$ is the mass of $^{40}$K, and $C_\beta$ is predicted to equal 10.1~\cite{Helfrich10}. The solid line in Fig.~\ref{fig:beta} shows a fit of Eq.~(\ref{eqn:betashape}) for $a$$>$$2R_\text{vdW}$$=$$144~a_\text{0}$~\cite{ChinRMP}, which gives $a_*$=$230(10)~a_\text{0}$, $\eta*$=$0.26(3)$, and $C_\beta$=$3.2(2)$. The variation in $a_*$ when changing the minimum fitted $a$ is approximately $\pm30~a_0$.  The discrepancy of $C_\beta$ with theory may be due to finite temperature~\cite{Helfrich09}.
 \begin{figure}
 \includegraphics[width=0.5\textwidth]{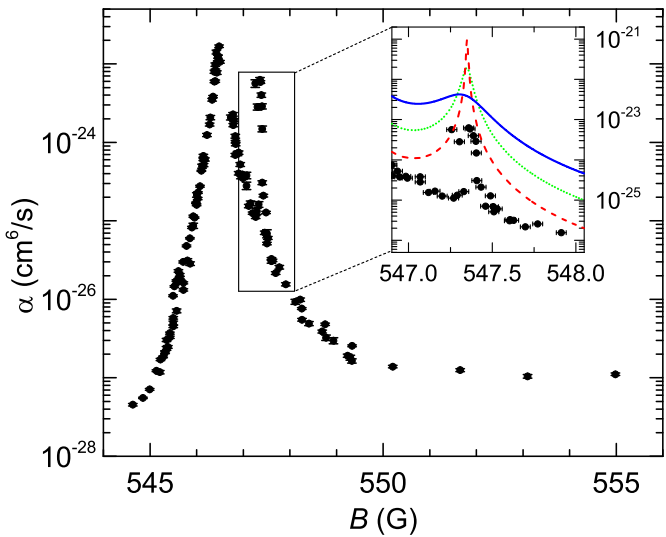}
 \caption{\label{fig:vsB}(Color online) $\alpha$ vs~$B$. There are two clear peaks near 546.6~G and 547.4~G: The first is the $s$-wave Feshbach resonance that we use to control $a$, and the second, much narrower feature we attribute to a $d$-wave interspecies Feshbach resonance.  Vertical error bars show the statistical uncertainty in $\alpha$, and horizontal error bars show an upper bound of $\pm0.03$~G for the variation in $B$ during the measurement. Inset: $\alpha$ vs.~$B$ near the feature at 547.4~G.  The solid, dotted, and dashed lines show the universal shape of an Efimov resonance, Eqs.~(\ref{eqn:alphaneg}) and (\ref{eqn:alphapos}), for $\eta*$=$0.1$, $\eta*$=$0.02$, and $\eta*$=$0.004$.}
 \end{figure}

We now turn our attention to three-body recombination, which is widely used to reveal Efimov physics in cold atoms. Three atoms collide inelastically and form a diatomic molecule, which typically releases enough energy to eject all three atoms from the trap. This process is resonantly enhanced when an Efimov trimer state approaches zero binding energy at a negative $a$ denoted $a_\text{-}$. On the other side of the Feshbach resonance, interference effects lead to minima in the three-body rate coefficient $\alpha$ at a positive $a$ denoted $a_\text{+}$.  These features modulate an overall $a^4$ dependence of $\alpha$ that results in a large increase of atom loss rates near a Feshbach resonance~\cite{Esry99,Braaten06}.

In the only previous observation of heteronuclear Efimov resonances, enhanced loss for $^{41}$K-$^{87}$Rb-$^{87}$Rb was observed at $a_-$=-246(14)~$a_\text{0}$~\cite{Barontini09}.   An additional narrow loss feature at 667(1)~$a_\text{0}$ indicated a possible atom-molecule resonance~\cite{Barontini09}.  In contrast, previous measurements of $\alpha$ for a $^{40}$K-$^{87}$Rb mixture saw no features in the $a^4$ scaling; however, the data were relatively sparse~\cite{Zirbel08}.  We present here additional measurements of $\alpha$ for $^{40}$K-$^{87}$Rb that cover values of $a$ where the $^{41}$K-$^{87}$Rb features were seen, in order to test the universality of the three-body parameter~\cite{WangY12b}.

The atom gas mixture is prepared by evaporating at a field where $|a|$$\approx$300~$a_\text{0}$ on the same side of the Feshbach resonance as the field for the loss measurement, $B$. The field is then swept to $B$ in a short time compared to the loss, and held for a variable time $t$.  Finally, the optical trap is turned off and the field is swept back to the evaporation value where the atoms are imaged. Two channels exist for interspecies three-body recombination; however, in our case, only $^{87}$Rb+$^{87}$Rb+$^{40}$K supports Efimov resonances~\cite{Braaten06}, while $^{87}$Rb+$^{40}$K+$^{40}$K loss is suppressed by Fermi statistics. For the first channel, $\alpha$ is defined by
\begin{equation}
 \dot{N}_\text{Rb}(t)=2\dot{N}_\text{K}(t)=-2\alpha \int d^3\textbf{r}~n_\text{K}(\textbf{r},t)~n^2_\text{Rb}(\textbf{r},t),
 \label{eqn:alpha}
\end{equation}
where $n_\text{Rb}(\textbf{r},t)$ and $n_\text{K}(\textbf{r},t)$ are the densities of $^{87}$Rb and $^{40}$K, respectively. From our measured cloud sizes and atom numbers, we obtain $\alpha$ using an analysis similar to that described in Ref.~\cite{Ospelkaus06}, which uses the measured atom densities at each time to account for heating of the gas. We estimate the systematic uncertainty in $\alpha$ to be $\pm20\%$.

Figure~\ref{fig:vsB} shows the measured $\alpha$ vs.~$B$. We measure an increase of $\alpha$ by over four orders of magnitude near the $s$-wave resonance at $B_0$=546.618~G.  A second peak, on the order of 0.1~G wide, is evident near 547.4~G.  As illustrated in the inset to Fig.~\ref{fig:vsB}, this feature is not consistent with the expected shape of an Efimov resonance, and we identify it as a two-body $d$-wave Feshbach resonance~\cite{Zaccanti06,Ruzic13}.  Because it is technically challenging to control $a$ near a narrow resonance, we exclude the data for $B$ between 547.25~G and 547.50~G from further analysis and calculate $a$ using the previously given equation for the $s$-wave resonance.

\begin{figure*}
\includegraphics[width=\textwidth]{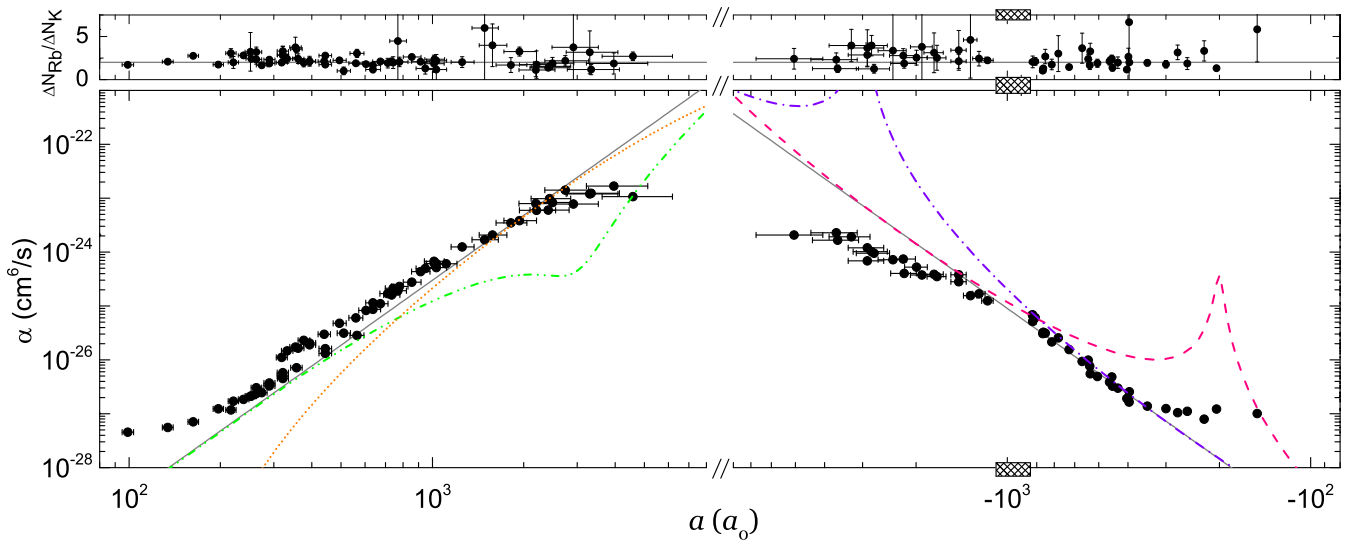}
 \caption{\label{fig:alpha}(Color online) Lower panel: $\alpha$ vs.~$a$. The error bars correspond to those in Fig.~\ref{fig:vsB}. The solid line shows an $a^4$ dependence with no Efimov resonances. To clarify the range of $a$ for which Efimov features are excluded, we include dashed and dash-dotted lines corresponding to universal Efimov resonance shapes located at $a_\text{-}$=$-200~a_\text{0}$ and $a_\text{-}$=$-3000~a_\text{0}$, respectively. Similarly, the dash-dot-dotted and dotted lines show hypothetical Efimov interference minima at $a_\text{+}$=$3000~a_\text{0}$ and $a_\text{+}$=$200~a_\text{0}$, respectively.  A narrow two-body Feshbach resonance blocks experimental access to the range $-1090$$\le$$a/a_\text{0}$$\le$$-840$ (shaded rectangles on axes). Upper panel: The ratio of $^{87}$Rb loss to $^{40}$K loss vs.~$a$. The solid line shows $\Delta\text{N}_\text{Rb}/\Delta\text{N}_\text{K}$=2.}
\end{figure*}

The upper panel of Fig.~\ref{fig:alpha} shows the measured ratio of $^{87}$Rb loss to $^{40}$K loss.  Within the uncertainties, this ratio does not show a dependence on $a$ and the average value is 2.1(1), which is consistent with $^{87}$Rb+$^{87}$Rb+$^{40}$K being the dominant loss process.  The lower panel of Fig.~\ref{fig:alpha} shows the measured $\alpha$ vs.~$a$.  These data are consistent with the more sparse measurements of Ref.~\cite{Zirbel08}. For comparison, we also plot the universal form for $\alpha$ as a function of $a$~\cite{Helfrich10} with Efimov resonances at hypothetical locations.  At negative $a$,
\begin{equation}
\alpha(a<0)=\frac{C_\alpha}{2}\frac{\text{coth}(\pi s_0)\text{sinh}(2\eta_*)}{\text{sin}^2[s_0~\text{ln}(a/a_-)]+\text{sinh}^2(\eta_*)}\frac{\hbar a^4}{m_1},
\label{eqn:alphaneg}
\end{equation}
whereas for positive $a$,
\begin{align}
\alpha(a&>0)=C_\alpha \bigg( \frac{\text{sin}^2[s_0~\text{ln}(a/a_+)]+\text{sinh}^2\eta_*}{\text{sinh}^2(\pi s_0 + \eta_*)+\text{cos}^2[s_0~\text{ln}(a/a_+)]}\nonumber\\
        &+\frac{\text{coth}(\pi s_0)\text{cosh}(\eta_*)\text{sinh}(\eta_*)}{\text{sinh}^2(\pi s_0 + \eta_*)+\text{cos}^2[s_0~\text{ln}(a/a_+)]} \bigg) \frac{\hbar a^4}{m_1},
\label{eqn:alphapos}
\end{align}
where the coefficient $C_\alpha$ is 345~\cite{Helfrich10}. We use $\eta *$=0.02 for the plotted curves, which gives the best agreement to the measured magnitude of $\alpha$.
We conclude that there are no Efimov features in $\alpha$ at negative $a$ between -200~$a_\text{0}$ and -3000~$a_\text{0}$ or at positive $a$ between 200~$a_\text{0}$ and 3000~$a_\text{0}$.

In the remainder of the paper, we discuss our results in the context of universal three-body predictions. For $^{87}$Rb+$^{87}$Rb+$^{40}$K, the predicted universal relationships for the location of Efimov features due to the same trimer state are $a_+/a_*$=22 and $|a_-|/a_*$=240 ~\cite{Helfrich10}. Given our observed resonance at $a_*$=$230(30)~a_\text{0}$ (Fig.~\ref{fig:beta}), these relations give $a_+$=$5100~a_\text{0}$ and $a_-$=$-55,000~a_\text{0}$. Scaling by $e^{\pi/s_0}$=122.7 for a more deeply bound Efimov state gives $a_+$=$41~a_\text{0}$ and $a_-$=$-450~a_\text{0}$.  The applicability of universal physics requires that $|a|$ be large compared to the physical range of the interactions, which can be characterized by $R_\text{vdW}$=$72~a_\text{0}$~\cite{ChinRMP}. For higher $|a|$, the useful range of $a$ is limited to $k|a|$$<$$1$, where $k$=$\sqrt{2\mu k_B T}/\hbar$ is the thermal wave number, $k_\text{B}$ is the Boltzmann constant, and $\mu$ is the reduced mass for $^{40}$K and $^{87}$Rb~\cite{DIncao04}. Our measurements of $\alpha$ at high $|a|$ are typically near $T$=$300$~nK, which gives $1/k$=$3000$~$a_\text{0}$. Within these bounds, the only observable Efimov feature in $\alpha$ would be at $a_-$=$-450~a_\text{0}$, which is excluded by our measurements.  However, the nonzero $^{87}$Rb-$^{87}$Rb scattering length could affect the relationships between resonance positions~\cite{WangY12b}. Additionally, even for the homonuclear case, the ratio of $|a_-|$ and $a_*$ measurements have not always agreed with universal Efimov theory~\cite{Knoop09, Wenz09, Lompe10, Berninger11}.

Wang \emph{et al.} predict universality of the three-body parameter for heteronuclear mixtures, and predict the locations of K+Rb+Rb Efimov resonances to be $a_+$=$2800~a_\text{0}$ and $a_-$$<$$-30000~a_\text{0}$~\cite{WangY12b}.  The corresponding atom-molecule loss feature would be at $a_*$=$130~a_\text{0}$. Our measurements of $\alpha$ exclude $a_+$=$2800~a_\text{0}$ and we find $a_*$=$230(30)~a_\text{0}$ from atom-molecule loss; both results suggest that the value of $a_+$ is higher than the prediction.

Finally, we compare our results for $^{40}$K-$^{87}$Rb to the previous measurements for $^{41}$K-$^{87}$Rb. The two mixtures share the same isotope of Rb, they both have resonant interspecies interactions with $R_\text{vdW}$=$72~a_\text{0}$~\cite{ChinRMP,Simoni08}, and their mass ratios differ by only a few percent. If the three-body parameter is universally determined by the two-body potential, one would expect $^{41}$K-$^{87}$Rb and $^{40}$K-$^{87}$Rb mixtures to have the same Efimov resonance locations~\cite{WangY12b}.  However, our measurements exclude $a_-$=-246~$a_\text{0}$, where the $^{41}$K-$^{87}$Rb-$^{87}$Rb Efimov resonance was observed. The $^{41}$K-$^{87}$Rb work also identified a feature in $\alpha$ at 667(1)~$a_\text{0}$ as a possible atom-molecule resonance~\cite{Barontini09}.  For the $^{40}$K-$^{87}$Rb mixture, we find $a_*$=$230(30)~a_\text{0}$ with direct measurements of atom-molecule loss and we do not observe a corresponding feature in $\alpha$, although we cannot rule out an extremely narrow resonance.

In the comparison of our results with theoretical predictions and with the data for a different isotopic mixture, we do not find clear evidence for a universal three-body parameter.  However, given the large Efimov scaling factor ($122.7$), the observed differences are relatively small. In addition, the clearest features seen for both mixtures occur at scattering lengths that are only a few times larger than $R_\text{vdW}$, where non-universal finite-range effects can play a role~\cite{Efimov91, Thogersen08, Platter09}.  This role may be greater for $^{40}$K-$^{87}$Rb, where the Feshbach resonance has an intermediate strength parameterized by $s_\text{res}$=$2.0$, compared to the strong $^{41}$K-$^{87}$Rb resonance with $s_\text{res}$=$26$~\cite{ChinRMP}. As was the case for homonuclear systems, data from additional heteronuclear atom mixtures will be important for elucidating the three-body parameter's universality as well as the role of finite-range corrections. In particular, heteronuclear systems with strong mass imbalance~\cite{Deh08,Tung13,Repp13} offer the best chance for observing multiple Efimov resonances at different $|a|$.

\begin{acknowledgments}
The authors thank Chris Greene, Jose D'Incao, John Bohn, Yujun Wang, and Eric Cornell for discussions. RSB acknowledges support from NSF-GRFP and NDSEG.  This work was funded by NIST and the NSF.
\end{acknowledgments}

\end{document}